\documentclass[epj]{svjour}
\usepackage{graphics}
\newcommand{\dzero}     {\ensuremath{\mathrm D^0}}

\newcommand{\dstplus}   {\ensuremath{\mathrm D^{*+}}}
\newcommand{\dststzero} {\ensuremath{\mathrm D^{**0}}}
\newcommand{\brbtodstst} {\ensuremath{\mathrm{BR (\bar{B} \rightarrow D^{**}\ell \bar{\nu} X)}}}
\newcommand{\bddx}      {\ensuremath{\mathrm{BR(b\rightarrow D \bar{D}X)}}}
\newcommand{\bdd}       {\ensuremath{\mathrm{b\rightarrow D \bar{D}X}}}

\newcommand{\bccx}      {\ensuremath{\mathrm{BR(b\rightarrow \psi X)}}}

\newcommand{\bnocharm}  {\ensuremath{\mathrm{BR(b \rightarrow 0 \; c)}}}
\newcommand{\lnpj}      {\ensuremath{\mathrm{-ln}(P_j)}}

\begin{document}
\title{Recent b-physics results from OPAL}
\author{David Waller\inst{1} 
}                     
\institute{Ottawa-Carleton Institute for Physics, Department of Physics, Carleton University, Ottawa, Canada}
\date{Received: date / Revised version: date}
%
\abstract{
The most recent b-physics results from the OPAL experiment at LEP are reviewed. A measurement of semileptonic B meson decays to narrow orbitally-excited charm mesons is presented first, followed by a study of charm production in b-hadron decays. Here, B refers to B$^+$ and B$^0$ mesons and their charge conjugates, and b-hadron refers to the admixture of hadrons containing a b quark produced in electron-positron annihilations at $\sqrt{s}= {\rm m_{Z}}$.
} 
\maketitle
\section{Introduction}
\label{intro}

Studying the decays of b-hadrons enables important tests of the Standard Model and Heavy Quark Effective Theory to be made. Two such tests are whether the observed branching ratios for semileptonic B decays and the average number of charm quarks produced per beauty quark decay are consistent with theoretical predictions. The predictions for these values are correlated due to their dependence on common physics inputs: the ratio of the charm and beauty quark masses, ${\rm m_c/m_b}$, and the renormalization scale, $\mu$, for the calculations of the relevant processes \cite{neubert}.  

Two recent OPAL results are presented in this paper. 
The first is a measurement of semileptonic B decays to narrow orbitally-excited charm mesons \cite{pedro} (in Section \ref{sec:semileptonic}) and the second is a study of charm production in b-hadron decays \cite{waller} (in Section \ref{sec:charm}).

\section{Semileptonic B decays to D$^{**0}$}
\label{sec:semileptonic}

Measuring the semileptonic decay of B mesons to narrow orbitally-excited charm mesons is of particular interest, as the sum of the measured exclusive semileptonic B decay branching ratios is less than the measured inclusive branching ratio \cite{richman,aleph,pdg}. Uncertainty in $\brbtodstst$ is also the largest source of uncertainty in measurements of $V_{\rm cb}$ at LEP, so it is desirable to measure this decay as precisely as possible.

The narrow orbitally-excited charm mesons are $L=1$ states where $J_q$, the total angular momentum of the light quark in the charm meson, is 3/2. In the heavy quark infinite mass limit, the light quark quantum numbers are conserved in heavy quark decays. The $J_q = 3/2$ states are narrow as they undergo D-wave decay. In this analysis, the two neutral narrow $\dststzero$ states are studied: ${\rm D^0_1}$ ($J^P = 1^+$) and ${\rm D^{*0}_2}$ ($J^P = 2^+$).

Instances of the decay chain 
\begin{eqnarray}
{\rm b \rightarrow \bar{B}} & \rightarrow & {\rm D^{**0} \ell \bar{\nu}X}, \nonumber  \\
{\rm D^{**0}} & \rightarrow & {\rm D^{*+} \pi^-}, \nonumber \\
{\rm D^{*+}}  & \rightarrow & {\rm D^{0} \pi^+_{\rm slow}},  \nonumber   \\
{\rm D^{0}}   & \rightarrow & {\rm (K\pi \; or \; K 3\pi)} \nonumber
\end{eqnarray}
are reconstructed in a sample of 3.9 million hadronic Z decays, collected from 1991-2000.

After the standard OPAL hadronic pre-selection, with an efficiency of (98.1$\pm$0.5)\% and a background of ($0.11\pm0.03$)\% \cite{hadron_sel}, events are divided into jets with a cone algorithm \cite{cone}. A high momentum lepton (e, $\mu$) candidate is then required in a jet. $\dzero$ candidates are reconstructed in the ${\rm K\pi \; and \; K 3\pi}$ channels from tracks in jets with identified high momentum leptons. If the $\dzero$ daughter tracks satisfy particle identification and invariant mass criteria, they are paired with slow pion candidates to reconstruct $\dstplus$ candidates. Pion candidates from $\dststzero$ decays are combined with the slow pion and lepton candidates to form a B vertex. If, in addition to passing a number of background suppression cuts, the decay length significances of the B with respect to the primary vertex and the D with respect to the B vertex are greater than 1.5 and -2.0 respectively, then the mass difference $\Delta \rm{m}^{**}$ between the $\dststzero$ and the $\dstplus$ is computed. 

\begin{figure}
\resizebox{0.5\textwidth}{!}{%
  \includegraphics{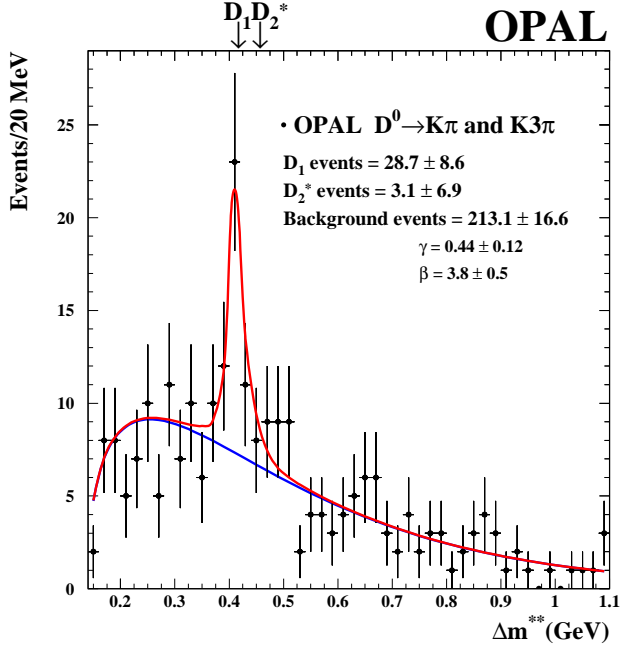}
}
\caption{Mass difference $\Delta \rm{m}^{**}$ between $\dststzero$ and $\dstplus$ candidates. The expected $\Delta \rm{m}^{**}$ values for the $\rm{D^{*0}_2}$ and $\rm{D^{0}_1}$ states are shown by the arrows at the top of the plot. The signal plus background and background alone shapes are shown by the lines superimposed on the data points. Shown on the plot are the best values for $\gamma$ and $\beta$, the terms used to parameterize the background.} 
\label{fig:mass_diff_fit}       
\end{figure}

An unbinned maximum likelihood fit of the $\Delta \rm{m}^{**}$ distribution is performed with a two-parameter background function and two Breit-Wigners, each convolved with a Gaussian, for the ${\rm D^0_1}$ and ${\rm D^{*0}_2}$ mass peaks (see Figure \ref{fig:mass_diff_fit}). The best fit values for the amplitudes of the signal peaks are used to determine the semileptonic B to $\dststzero$ branching ratios. The largest sources of systematic uncertainty in the analysis are the parameterizations of the signal and background functions. The sensitivity of the analysis to the shape of the background function is reduced by simultaneously fitting the ``right-sign'' data sample and a ``wrong-sign'' data sample, where a pion candidate with the wrong charge for the $\dststzero$ decay is selected instead of a right-sign pion.

The resulting fit gives the product branching ratios
${\rm BR(b \rightarrow \bar{B}) \times BR(\bar{B} \rightarrow D^0_1 \ell^- \bar{\nu}X) \times BR(D^0_1 \rightarrow \dstplus \pi^-)}$ \\ ${\rm= (2.64\pm0.79(stat.)\pm0.39(syst.)) \times 10^{-3}}$,
and \\
${\rm BR(b \rightarrow \bar{B}) \times BR(\bar{B} \rightarrow D^{*0}_2 \ell^- \bar{\nu}X) \times BR(D^{*0}_2 \rightarrow \dstplus \pi^-)}$ \\ ${\rm< 1.4 \times 10^{-3} \; at \; 95\% \; C.L.}$.
These results update an earlier OPAL result \cite{opal_old} and are consistent with similar results from ARGUS \cite{argus}, CLEO\cite{cleo}, and other LEP experiments\cite{aleph,delphi}.

\section{Charm counting in b-hadron decays}
\label{sec:charm}

The average number of charm plus anti-charm quarks produced per beauty quark decay, $n_c$, is predicted to be $1.20 \pm 0.06$ \cite{neubert}. In this analysis, an inclusive method that differentiates between different b-hadron decay topologies is used to measure $\bddx$\footnote{Here, D refers to any charm hadron.}. This branching ratio is used to calculate $n_c$:
\begin{equation}
n_c \! = \! 1 \! + \! \bddx \! + \! \bccx \! - \! \bnocharm,
\label{eqn:nc}
\end{equation}
where $\bccx$ is the inclusive branching ratio to charmonium and $\bnocharm$ is the inclusive branching ratio to final states containing no c or ${\rm \bar{c}}$ quarks.

Data collected from 1993 to 1995 are used for this analysis.  After the same hadronic pre-selection mentioned in Section \ref{sec:semileptonic}, plus additional cuts on the event thrust and thrust axis, a sample of 1.9 million hadronic Z decays remains. A high purity and efficiency b-tagging algorithm is applied to the data to obtain a 95\% pure sample of b-jets \cite{opal_higgs_btag}.  Jets opposite tagged b-jets are used in the analysis to avoid bias in the selection of different b-hadron decay topologies. The impact parameter significances $S$ of all tracks that pass strict selection cuts in a jet are used to calculate a single joint probability variable $P_j$ for discriminating $\bdd$ decays from other b-hadron decays and backgrounds. Jets containing tracks with large $S$ values (e.g. tracks from $\bdd$ decays) tend to have small $P_j$ values.

\begin{figure}
\resizebox{0.5\textwidth}{!}{%
  \includegraphics{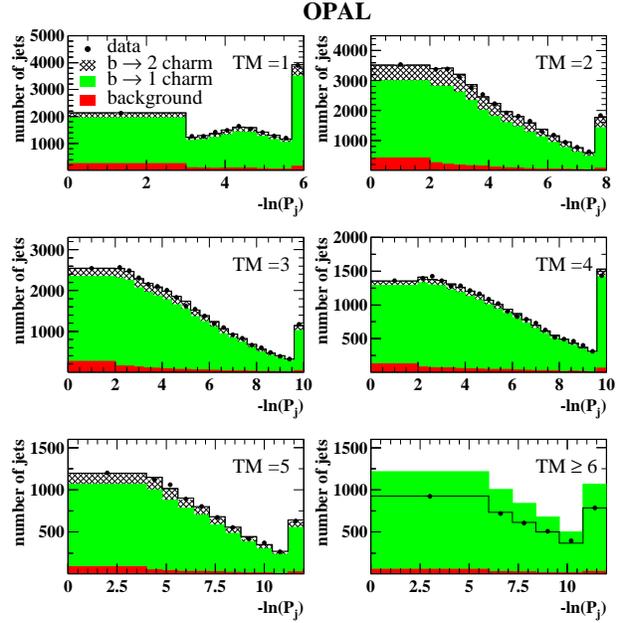}
}
\caption{Sums of fitted Monte Carlo $\lnpj$ distributions (solid line) and data (solid points) for each track multiplicity, TM, bin.  The shaded region at the top/bottom of each plot represents the $\bdd$/background component. The sum of the Monte Carlo distributions is smaller than the ${\rm b \rightarrow 1 \; charm}$ component for TM$\geq6$ as the best fit number of $\bdd$ jets is negative.} 
\label{fig:charm_fit}       
\end{figure}

The $\lnpj$ distributions of Monte Carlo simulations for different b-hadron decays and backgrounds are compared to data in order to determine $\bddx$. The $\lnpj$ distributions are binned according to the multiplicity of tracks contributing to the calculation of $\lnpj$ for each jet. A $\chi^2$-fit is performed to determine the number of $\bdd$ and single charm b-hadron decays in the data; the branching ratios of b-hadrons to final states with no open charm are fixed at the current world average values \cite{pdg}, and the fraction of non-b-jets that pass the b-tag is fixed to a value determined from Monte Carlo studies. A small additional term, $\alpha$, (of order $10^{-3}$) is included in the fitting function in order to reduce the sensitivity of the analysis to many sources of systematic uncertainty. The inclusion of this term allows the data to constrain the physics inputs that change the $\lnpj$ distributions in an a linear manner. A zero value for $\alpha$ implies perfect modelling.  Due to the excellent modelling of the physics processes and the detector response, the best fit value for $\alpha$ is consistent with zero.  Figure \ref{fig:charm_fit} shows the result of the fit.

The sources of the largest systematic uncertainties in the analysis are the number of tracks from fragmentation ({\it i.e.} tracks that are not the products of b-hadron decays) and the number of neutral pions produced in charm hadron decays. Varying the values of these quantities in the Monte Carlo significantly affects the shapes of the simulated $\lnpj$ distributions. 

The result obtained in this analysis is 
 $\bddx = (10.0 \pm 3.2 ({\rm stat.})^{+2.4}_{-2.9}({\rm det.})^{+10.4}_{-9.0}({\rm phys.}))\%$,  
where ``det.'' and ``phys.'' refer to the uncertainties due to detector and physics modelling, respectively.  This branching ratio is input to equation \ref{eqn:nc} to yield $n_c = 1.12^{+0.11}_{-0.10}$.  These values of $\bddx$ and $n_c$ are consistent with the corresponding experimental averages \cite{pdg,lephfwg} and theoretical predictions \cite{neubert}. The systematic uncertainty due to physics modelling is significantly larger than the uncertainty quoted for a similar DELPHI analysis \cite{delphi_nc}; however, the sources of the largest uncertainties in this analysis were not assigned values in the DELPHI analysis. 

\section{Conclusion}

Two recently completed b-physics analyses from the OPAL collaboration have been reviewed in this paper. The first is a measurement of semileptonic B meson decays to narrow orbitally-excited charm mesons. The second is a study of charm production in b-hadron decays. The results of both analyses agree with theoretical expectations and the results from other experiments.

%
%


\end{document}